# Coupled-Oscillator Associative Memory Array Operation

Dmitri E. Nikonov [1], Gyorgy Csaba [2], Wolfgang Porod [2], Tadashi Shibata [3], Danny Voils [4], Dan Hammerstrom [4], Ian A. Young [1], and George I. Bourianoff [1]

1 Components Research, Intel Corp., Hillsboro, Oregon 97124 USA

2 Center for Nano Science and Technology, University of Notre Dame, Notre Dame, Indiana 46556, USA

3 University of Tokyo, Tokyo, Japan

4 Portland State University, Portland, Oregon

Abstract

Operation of the array of coupled oscillators underlying the associative memory function is demonstrated for various interconnection schemes (cross-connect, star phase keying and star frequency keying) and various physical implementation of oscillators (van der Pol, phase-locked loop, spin torque). The speed of synchronization of oscillators and the evolution of the degree of matching is studied as a function of device parameters. The dependence of errors in association on the number of the memorized patterns and the distance between the test and the memorized pattern is determined for Palm, Furber and Hopfield association algorithms.

# Contents

## 1. Introduction

The progress of integrated circuits for digital computing has been an unprecedented success for the past 40 years due to scaling of the number of transistors on chip (Moore's law [1]). Continued scaling is projected for at least another decade [2]. Digital circuits thus handily meet user requirements for processing of numerical, text and video information. However there is a class of problems, traditionally associated with human intelligence, that computers do not handle as successfully. They are, for example, image recognition, speech recognition, contextual search, detection of spatio-temporal events, etc. Algorithms for their solution based on digital Boolean logic do exist, but require excessive computational effort. Various researchers arrived at the idea to explore alternative analog or non-Boolean methods of computing for these problems. A school of thought emerged that aimed to emulate, to various degrees, operation of neurons. It resulted in vigorous growth of the fields of neural networks [3] and neuromorphic computing [4].

Various architectures for non-Boolean computing exist. Artificial neural networks [5] are cascaded devices with typically high fan-in and fan-out. Cellular neural networks [6] are typically rectangular arrays of nodes, each connected to nearest neighbors. LEGION networks [7] combine coupling between nearest neighbor oscillators with a common inhibiting node. In contrast to the above approaches we are dealing in this paper with networks of oscillators which we call coupled oscillator associative memory array (COAMA). In such a network all oscillators are coupled to each other, possibly through a common node (averager). The memorized and input patterns are encoded as parameters of oscillators. Under proper operation, if an input pattern is close to one of the memorized patterns, the phases of oscillators synchronize and we interpret it as recognition. However if the phases of oscillators do not synchronize, we talk about lack of recognition.

Our work builds on prior research by Hoppensteadt and Izhikevich [8], Itoh and Chua [9], Corinto, Bonnini, and Gilli [10]. But in this report we are going further. We design realistic schemes of such oscillator arrays using particular nanoscale devices (such as nanotransistors and spin torque oscillators).

Prior work was using the scheme in which the patterns are encoded as constants of coupling between oscillators (phase shift keying, PSK). Correspondingly, two stages – initialization and recognition were required in PSK. Here we present for the first time a scheme of frequency shift keying, FSK, in which patterns are encoded as changes of frequencies of oscillators. FSK requires only a single stage of recognition. We use more realistic mathematical models for simulating oscillators. Specifically, the Kuramoto model [11] has been widely used to represent arrays of coupled oscillators. It contains only phases of oscillators. In contrast, all our models involve both amplitudes and phases of oscillators. This holds true for our phase-locked loop model being more rigorous than that in in Ref. [12]. For spin torque oscillators we extend our treatment from the macrospin model (describing the magnetization of a nanomagnet by a single vector) to a micromagnetic simulation (capturing the coordinate dependence of magnetization).

For realizing dynamic, non-Boolean computing systems we explore an avenue where basic device components are not trying to imitate a CMOS switch or circuit dynamics, but where individual device components itself are complex dynamical systems. The benefits of oscillatory non-Boolean systems could potentially be better exploited using such devices. As an example, we study Spin-Torque Oscillators (STO) in Ref. [13]. STOs are sub 100 nanometer-scale devices, acting as compact microwave oscillators [14,15,16,17,18,19]. The self-sustaining oscillations are generated by the flow of spin-polarized currents (spin-torque) into a thin magnetic layer and the magnetization oscillations can be detected by the resistance change in the same current path [20]. The magnetic oscillations may create propagating spin waves [21,22], which provide a non-electrical interaction mechanism between STOs [23].

The physics underlying STO operation is, in principle, well understood and one can use standard micromagnetic simulation codes [24] to model magnetization dynamics. However, due to the strong nonlinearity of magnetization dynamics [25], the large variety of possible oscillation modes [26] and spin-wave modes and the heavy computational workload of a full micromagnetic simulation, one has to use a hierarchical set of approximate modeling tools to understand the behavior of spin torque oscillator networks. This paper presents such modeling hierarchy.

The expressions of coupling between oscillators map on various algorithms of sparse data representation in associative memories (Palm, Furber/Willshaw, or Hopfield; see Section 6). In this work we analyze the accuracy of recognition of these algorithms, first using random data patterns and then feature patterns extracted from real life images.

The paper is organized as follows. In Section 2, we overview the principle of an associative memory and a mathematical model for COAMA using an example of simple nonlinear oscillators. Data encoding by phase shift keying and frequency shift keying as well as the resulting synchronization of oscillators are presented in Section 3. Associative memory operation based on phase-locked loops is shown in Section 4, while based on spin torque oscillators – in Section 5. Recognition accuracy and information gain and their dependence on the sparse representation of data in an associative memory are treated in Section 6.

## 2. Associative memory and oscillator arrays

The role an associative memory is to compare a vector of a test pattern to the set of vectors of memorized patterns and to find one (or several) closest according to some metric defined for these vectors:

$$test = \xi_0, memorized = \xi_1, \xi_2, \xi_3, \qquad (1)$$

where the vectors can be of any length, binary or grayscale. A simple example of patterns in Figure 1 will be used for illustration here.

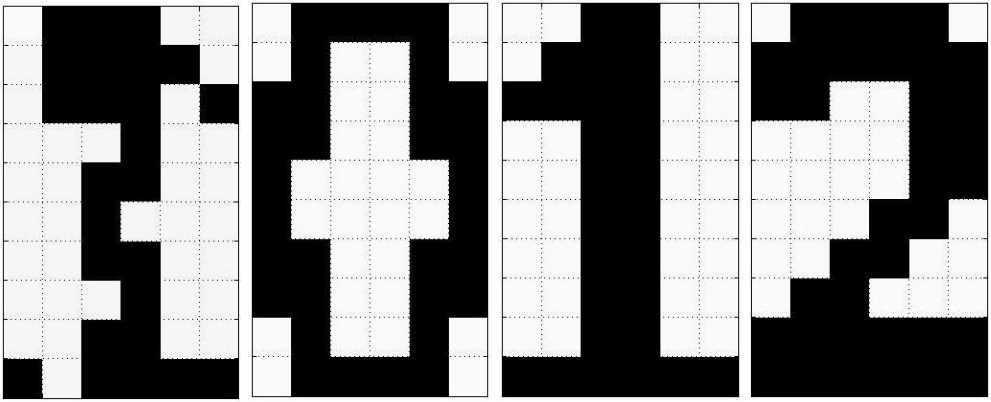

**Figure 1. Example 1x60 patterns. Leftmost = test pattern, 3 on the right = memorized pattern. The "1"-looking test pattern is the closest to the middle memorized pattern.**

We apply an array of coupled oscillators to the task of recognition and seek the design where synchronization of oscillators would correspond to a match. We start with describing the evolution of an array of nonlinear oscillators (complex van der Pol oscillators) by the following equations

$$\frac{dz_i}{dt} = (\rho_i + i\omega_i)z_i - z_i|z_i|^2 + \varepsilon\sum_{j=1}^{n} C_{ij}z_j, \qquad (2)$$

where $z = x + iy$ is the complex amplitude of an oscillator, $\rho_i$ is a parameter determining the limiting cycle amplitude, $\varepsilon$ is the strength of coupling between oscillators. This model closely follows multiple publications, e.g. [8,9,10]. In these examples we will use $\rho_i = 0.03$ and $\varepsilon = 0.01$ unless stated otherwise.

We consider two methods of encoding of patterns into oscillators: frequency- and phase-shift keying. We are not aware of prior publications of the first scheme. The mathematics of the second scheme has been described in [8,9,10].

In the frequency shift keying (FSK), the patterns are encoded as the frequency shifts of the oscillators. Each associative array compares the test vector with index *0* to one memorized vector with index *m*:

$$\omega_i = \omega_0 + \Delta\omega\left(\xi_{0,i} - \xi_{m,i}\right). \tag{3}$$

The coupling constants are set to a fixed value, e.g.

$$C_{ij} = 1. \tag{4}$$

In this case the degree of matching between patterns is calculated as

$$d_m = \frac{1}{n}\left|\sum_{i=1}^{n} z_i\right|, \tag{5}$$

and corresponds to the amplitude of the signal at the averager. The block-diagram of the FSK implementation is shown in Figure 2.

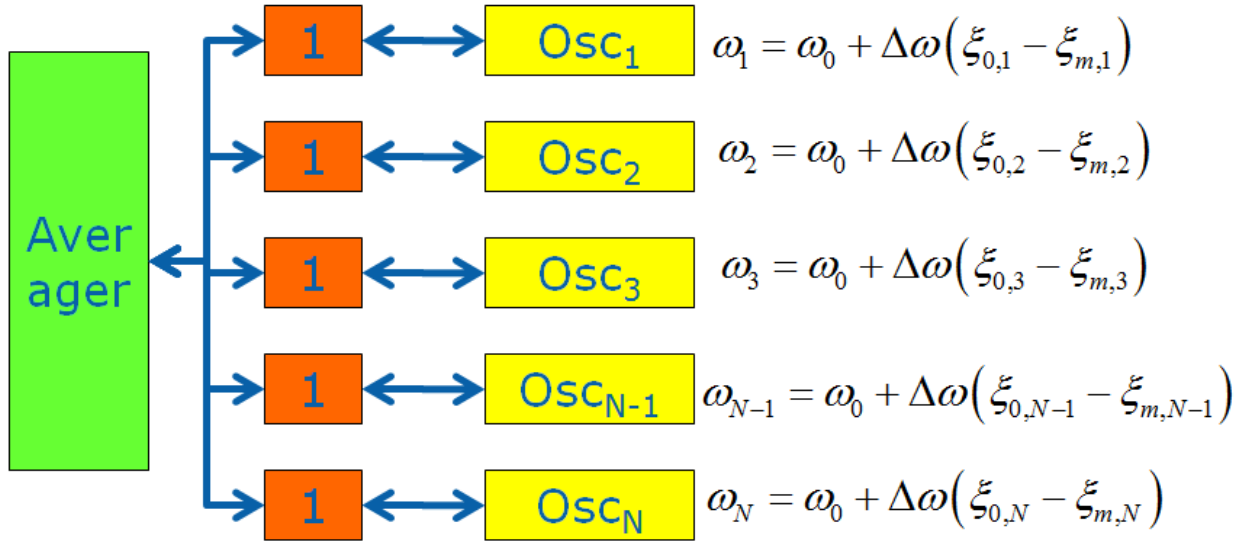

**Figure 2. Block diagram of an associative array in the FSK method. Signals are passed in both directions through links and multiplied by a factor each time passing an orange box.**

In the case of phase-shift keying (PSK), the following mapping of patterns on physical values is used: the logical bit value $b$ takes values of 1 or 0, and corresponds to the phase of oscillators $\phi = \pi b$, and the pattern values are $\xi = \cos\phi$ and take values from 1 to -1 and, respectively from white to black in the gray scale. The center frequencies of all oscillators are set to a fixed value $\omega_0$. The coupling constants are set by the product of pattern values. For the first stage, initialization, the pattern values are determined by the test vector $\xi_0$, such that

$$C_{ij}(init) = \xi_{0,i}\xi_{0,j}^{\dagger}. \tag{6}$$

The purpose of the initialization stage is to impose the phase differences corresponding to the test pattern on the array of oscillators starting from random initial conditions.

For the second stage, recognition, the coupling constants are switched to the ones determined by all the memorized vectors

$$C_{ij}(recog) = \frac{1}{m}\sum_{k=1}^{m}\xi_{k,i}\xi_{k,j}^{\dagger}. \tag{7}$$

In other words, all of the memorized patterns participate in the determination of dynamics of oscillators. Experience shows [8] that each pattern corresponds to a trajectory in the configuration space. The purpose of the recognition stage is, for thus prepared oscillators, to transition to the phase differences corresponding to one memorized pattern closest to the test pattern. In both stages, the degree of matching of the oscillator state to any vector with index $k$ is given by

$$d_k = \frac{1}{n}\left|\sum_{i=1}^{n}\xi_{k,i}z_i\right|. \tag{8}$$

The PSK can be implemented with two topologies: star [9,10] or cross-connect [8]. Both are mathematically equivalent (if one does not take into account realistic details of implementations, such as attenuation and delay of signals) and thus would produce identical simulation results.

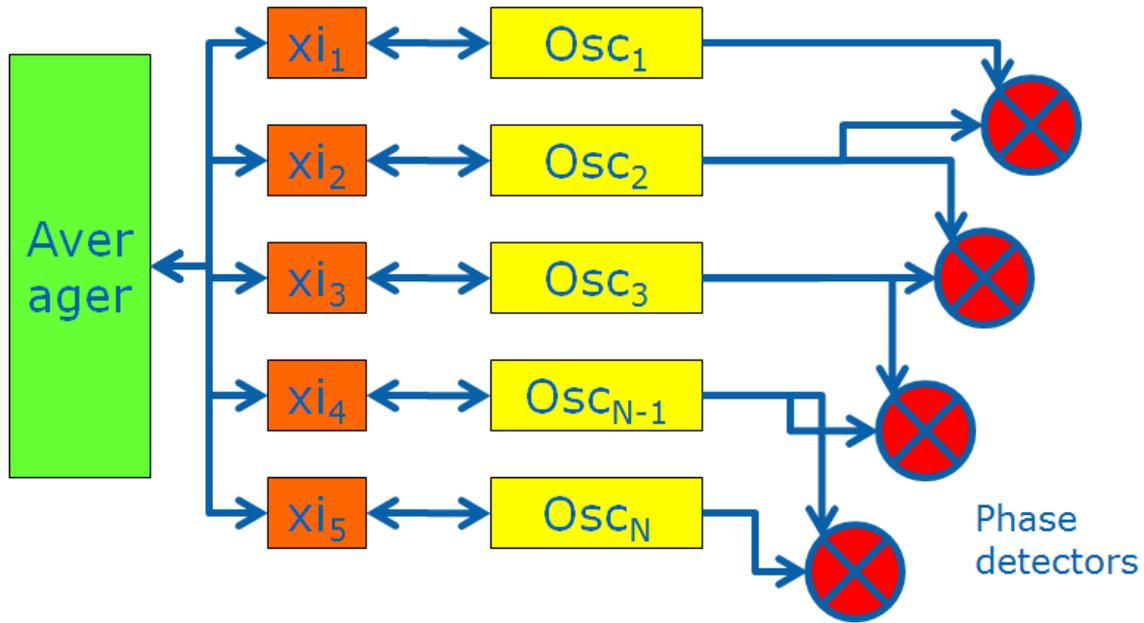

**Figure 3. Block-diagram of an associative array in the PSK method with star connection topology in the initialization stage. Signals are passed in both directions through links and multiplied by a factor each time passing an orange box. Phase detectors provide relative phases of neighboring oscillators.**

For the initialization stage, the oscillators are connected to one averager, Figure 3. For the recognition stage, the oscillators need to be disconnected from the averager corresponding to the test pattern and connected to a set of averagers corresponding to the memorized patterns, Figure 4. For *M* memorized patterns one need *M* averagers.

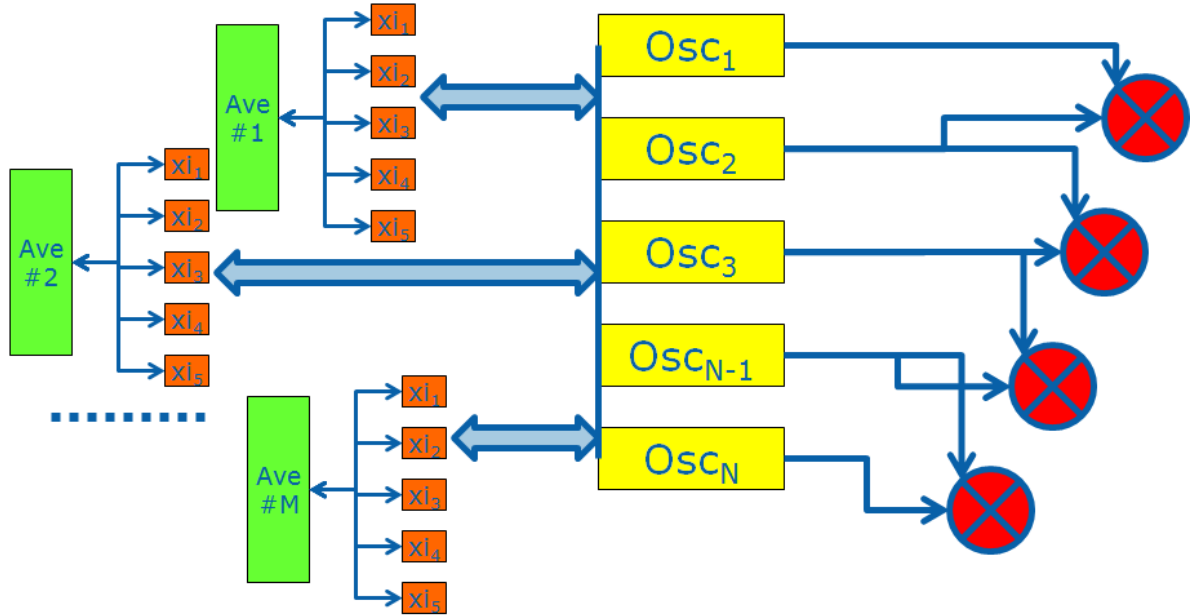

**Figure 4. Block-diagram of an associative array in the PSK method with star connection topology in the recognition stage. Similar to that of Figure 3, but the array of oscillators is simultaneously connected to M averages.**

In the cross-connect implementation, Figure 5, there are no averagers, and the degree of matching cannot be directly obtained. Thus one is forced to determine the differences of the phases between neighboring oscillators and compare them to the memorized patterns. This puts this implementation at a disadvantage compared to the star architecture. Its advantage is that for the recognition stage, the scheme is the same and the same number of coupling elements is needed for any number of memorized patterns.

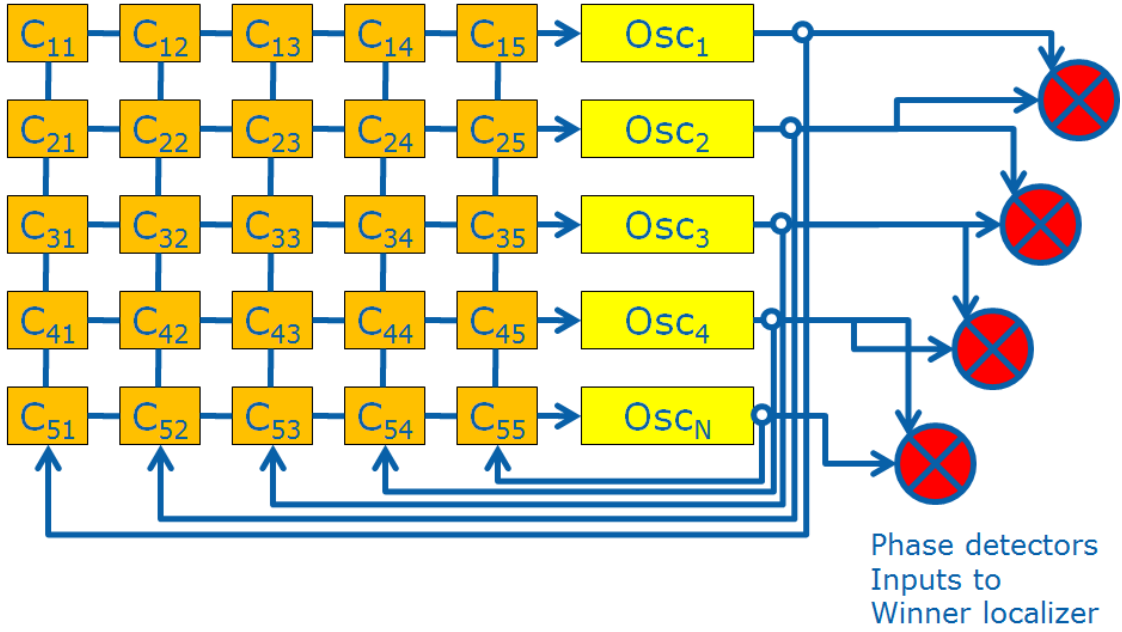

**Figure 5. Block-diagram of an associative array in the PSK method with cross-connect topology in either the initialization or recognition stages. Signals are passed in one direction in a loop and multiplied by a factor C each time passing an orange box. All signals in rows are summed and sent to drive inputs of oscillators. Phase detectors provide relative phases of neighboring oscillators.**

3. Phase shift keying vs. frequency shift keying synchronization

The results of simulation of arrays with randomly set initial conditions are shown below. Plotted are the phase differences of each of the 60 oscillators and the first oscillator. Time is in units of inverse cyclic central frequency of the oscillators $\omega_0^{-1}$. We see that for the case of good match, Figure 6, the phases converge to constant values different by $2\pi$ from each other. This proves that the oscillators are running at the same frequency and moreover are phase locked (synchronized). The degree of match reaches a high value (close to 1) and oscillates weakly around it. Thus synchronization is achieved over 6-10 periods of oscillation. Conversely, for the case of bad match, Figure 7, the phases of most oscillators continue to increase linearly. This indicates that synchronization has not occurred for some of the oscillators and the oscillators are running independently with their own frequencies. The degree of matching oscillates with a large amplitude around a small value. These features of the degree of match allow one to build circuits for determination of a winner-take-all (WTA) or k-WTA.

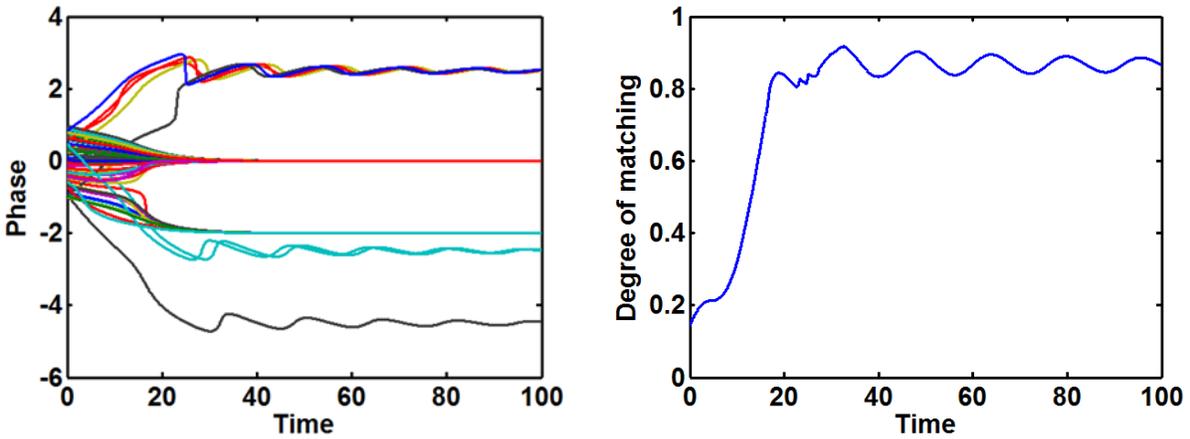

**Figure 6. Relative phases of oscillators in units of π (left) and the degree of match (right) vs. time for the associative array comparing the test pattern and the "1"-looking memorized pattern.**

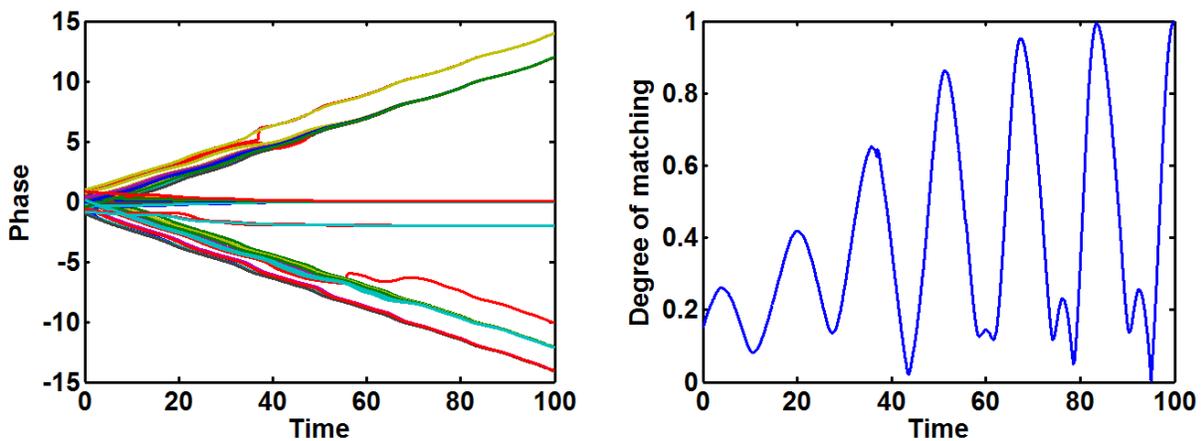

**Figure 7. Relative phases of oscillators in units of π (left) and the degree of match (right) vs. time for the associative array comparing the test pattern and the "0"-looking memorized pattern.**

The results of simulations of the evolution of oscillators in their initialization stage are in Figure 8, and for the recognition stage are in Figure 9. They show that oscillators, starting from random initial amplitude and phase, quickly converge to phases different by π or 2π, depending on the sign of coupling constants $C$. The degree of match to the test vector reaches 1. Then only a few oscillators switch their phases by p until the degree of match to one memorized vector increases to 1 while to others decreases to 0. The overall comparison of the FSK and PSK methods is given in Figure 10.

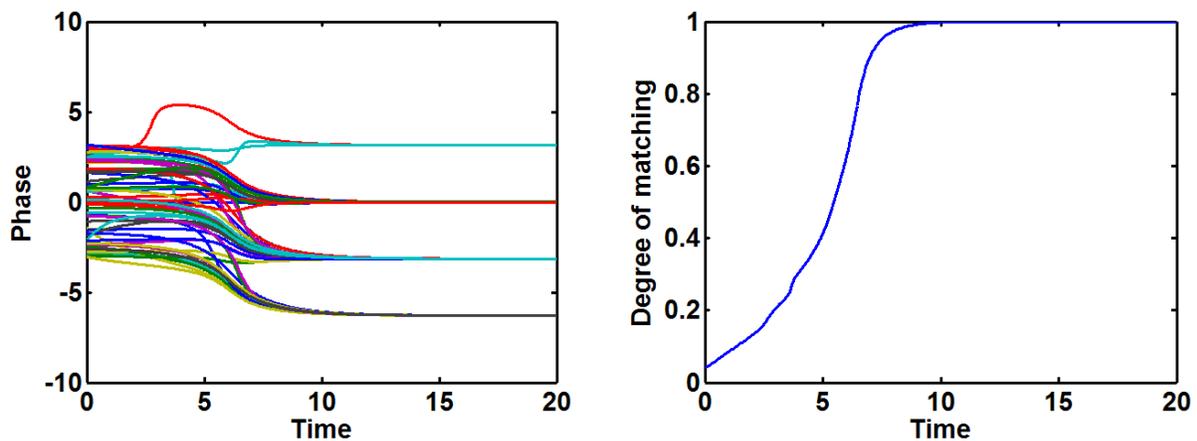

**Figure 8. Relative phases of oscillators (left) and the degree of match (right) vs. time for the associative array in the initialization stage.**

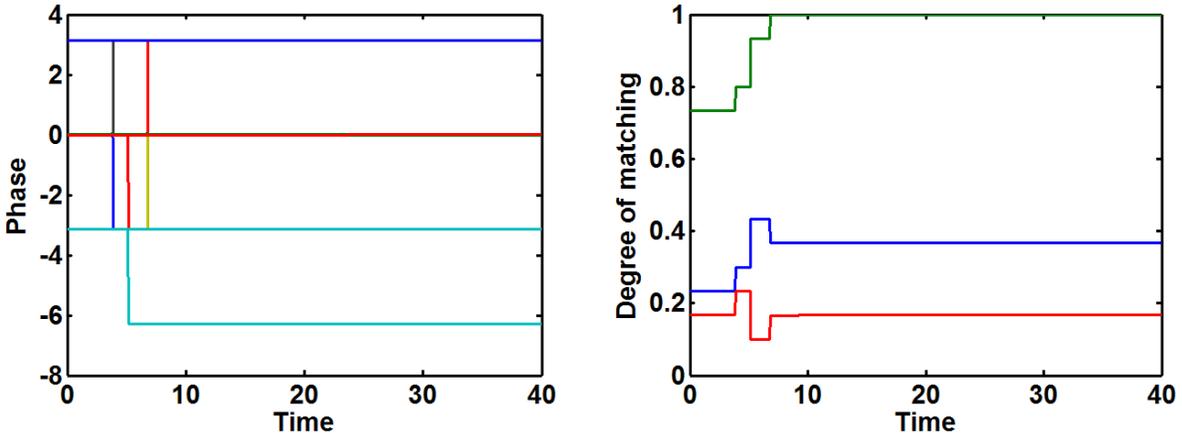

Figure 9. Relative phases of oscillators (left) and the degree of match (right) vs. time for the associative array in the recognition stage.

| Attribute | Phase keying | Frequency keying |
|---|---|---|
| Memorized patterns | M | 1 |
| Averagers (op-amp) | M | 1 |
| Oscillators | N | N |
| Memory elements | M*N | 0 |
| Coupling | Phase shift | Constant |
| Grey-scale inputs | Cannot | Can |

Figure 10. Table of comparison of characteristics of FSK and PSK associative arrays.

## 4. Phase locked loop synchronization

Another popular implementation of oscillators is a phase-locked loop (PLL) as shown in Figure 11.

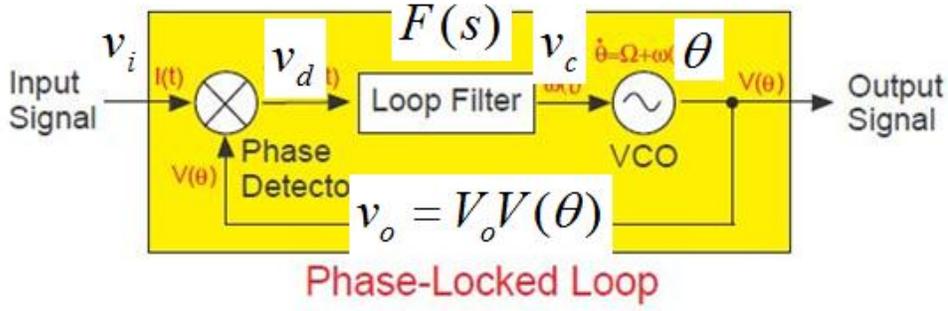

**Figure 11. Block diagram of a single phase-locked loop. It represents one oscillator in the array block diagrams.**

Here we provide a more general and rigorous model than in [12]. Each wire carries a sum of harmonic signals with a certain amplitude and phase. For the voltage-controlled oscillator (VCO) its phase derivative is modified by the input voltage

$$\frac{d\theta}{dt} = \omega + K_0 v_c, \qquad (9)$$

where $K_0$ is the gain of a VCO. Its input voltage is the output of a loop filter, described here by a one-pole transfer characteristic

$$\tau \frac{dv_c}{dt} = -v_c + v_d, \qquad (10)$$

And in turn the input signal is a coming from a phase detector which can be implemented as an ideal mixer with a factor $A_m$

$$v_d = A_m v_{in} v_o. \qquad (11)$$

We assume that the waveform generated by the oscillator is simple harmonic

$$V(\theta) = \cos(\theta). \qquad (12)$$

Then the equations for an array of linearly coupled VCOs are

$$\frac{d\theta_i}{dt} = \omega_i + K_0 v_{c,i}, \tag{13}$$

$$\tau \frac{dv_{c,i}}{dt} = -v_{c,i} + 2\varepsilon K_d V(\theta_i) \sum_{j=1}^{n} C_{ij} V(\theta_j - \pi/2), \tag{14}$$

where the gain of the mixer is

$$K_d = \frac{A_m V_o^2}{2}. \tag{15}$$

The degree of matching is determined in a manner similar to above:

$$d_k = \frac{1}{n} \left| \sum_{i=1}^{n} \xi_{k,i} \exp(i\theta_i) \right|. \tag{16}$$

The simulation results, Figure 12 and Figure 13, are qualitatively similar to the case of nonlinear oscillators. This supports the argument that the operation of an associative array is insensitive to the nature of an oscillator and the degree of its nonlinearity.

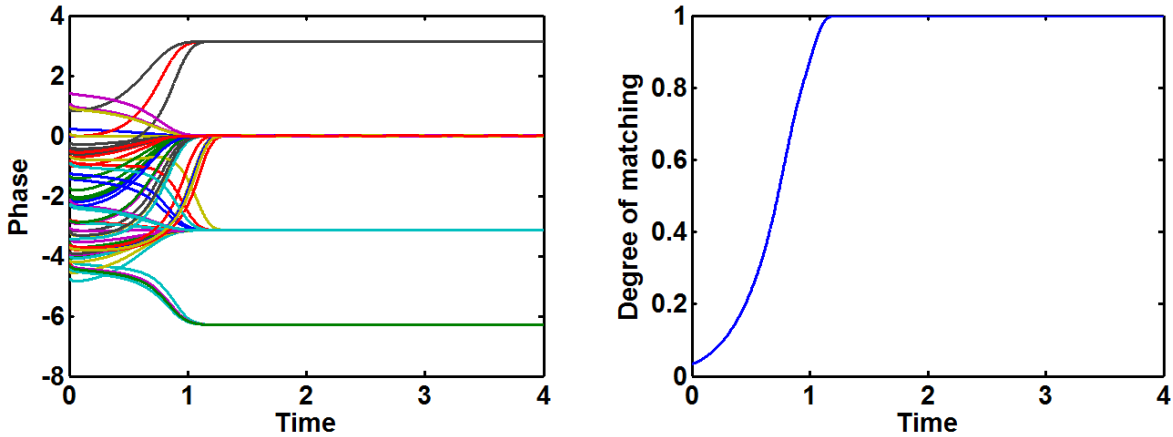

**Figure 12. Relative phases of PLLs (left) and the degree of match (right) vs. time for the associative array in the initialization stage.**

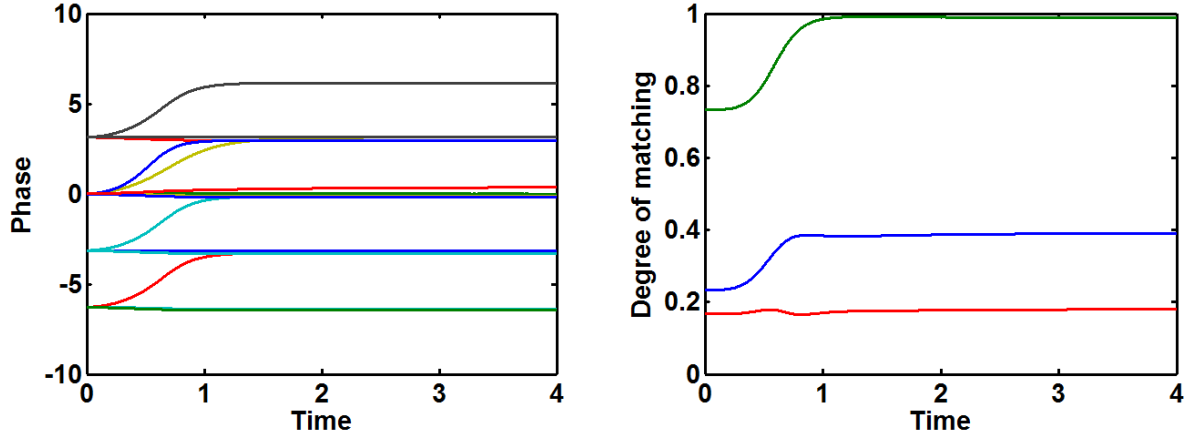

**Figure 13. Relative phases of PLLs (left) and the degree of match (right) vs. time for the associative array in the recognition stage.**

## 5. Spin torque oscillator synchronization

Spin torque oscillators (STO) are promising components for oscillatory associative memories. They dissipate very little power (can be driven by milliamperes of currents and millivolts of voltage), produce oscillations in the GHz range (which is highly compatible with microelectronic circuitry) and enable various interconnection topologies (passive and electrical connections as well as direct spin-wave coupling) [16]. Phase locking of two STO has been demonstrated [23].

The dynamics of STOs is more complicated than that of an idealized oscillator. It should be investigated whether they show synchronization behavior similar to Kuramoto phase oscillators [11] or the previously described van der Pol non-linear oscillators. To this end we describe how: 1) the partial differential equations (PDEs) micromagnetic equations describing spin-torque driven magnetization dynamics can be written in the forms of ordinary differential equations (ODEs); 2) how phase and frequency locking develops in systems of coupled STOs.

**A. Micromagnetic modeling**

Numerical solution of the Landau-Lifshitz equation (LLG) is widely used for physics-based simulation studies of the magnetization dynamics in sub-micrometer-size nanomagnets. In order to model spin-torque effects, the LLG equations are complemented with the Slonczewski spin-torque term [16]:

$$\left.\frac{d\mathbf{M}}{dt}\right|_{prec} = -\gamma(\mathbf{M}\times\mathbf{H}_{eff}) + \frac{\alpha\gamma}{M_s}[\mathbf{M}\times(\mathbf{M}\times\mathbf{H}_{eff})] + \frac{\gamma\beta\varepsilon}{M_s}[\mathbf{M}\times(\mathbf{S}\times\mathbf{M})]$$
$$\beta = \left|\frac{\hbar}{\mu_0 e}\right|\frac{J}{tM_s} \quad \varepsilon = \frac{P\Lambda^2}{(\Lambda^2+1)+(\Lambda^2-1)(\mathbf{M}\cdot\mathbf{S})} \quad (17)$$

Where $\mathbf{M}$ is the magnetization vector distribution of the free magnetic layer, $g = 2.210\times10^5$ m/(As) is the Landau-Lifshitz gyromagnetic ratio, $e$ is electron charge, $\mathbf{H}_{eff}$ is the effective magnetic field (which includes contributions from the STO shape, anisotropy and exchange stiffness), $M_s$ is the saturation magnetization (we used $M_s = 8.6\times10^5$ A/m in all simulations), α is the damping constant, $J$ is the current density, Λ is the spin asymmetry parameter (we used Λ=1.5 everywhere), $t = 5$ nm is the thickness of the free layer and $\mathbf{S}$ is a unit vector indicating the spin polarization of the driving current. The vector $\mathbf{S}$ is determined by the magnetization direction of a polarizer magnetic layer, but the magnetization of this layer is assumed to be fixed and not simulated. The above parameters are similar to STOs discussed in [19].

These equations can be applied in two different ways. As written above, they are Partial Differential Equations (PDEs), which give the response of a magnetization distribution to an applied external field and current distribution – these are all vector field variables. Micromagnetic solver packages, such as the well-established OOMMF code [24] are available for the solution, but solving these PDEs is time consuming, and it is difficult to find connections to a Kuramoto-type phase model.

For sufficiently small-sized STOs, one can replace the magnetization distribution with a magnetization vector and solve for only ordinary differential equations (ODEs) instead of PDEs. This approximation is usually referred as the single-domain or macrospin model. The **M** and **H** vector fields are represented by their volume average over the free layer. There is no general rule when such approximation is valid, so the macrospin model should always be carefully validated against the full micromagnetic model.

We performed this comparison for various-sized STOs. For an STO with an $d < 30$ nm diameter free layer, the macrospin model yields almost identical results to the PDE-based, full micromagnetic description. The frequency-current plot of Figure 14 gives a side-by-side comparison. For contact diameters between 30 nm and 50 nm, the macrospin model is a close approximation, but the full micromagnetic model predicts higher oscillation bandwidth, due to non-uniformities appearing in $\mathbf{M}(\mathbf{r},t)$. For contact diameters above 50 nm, the free layer magnetization breaks up into multiple domains during oscillation, and the linewidths, threshold and cut-off currents predicted by the macrospin model became highly inaccurate.

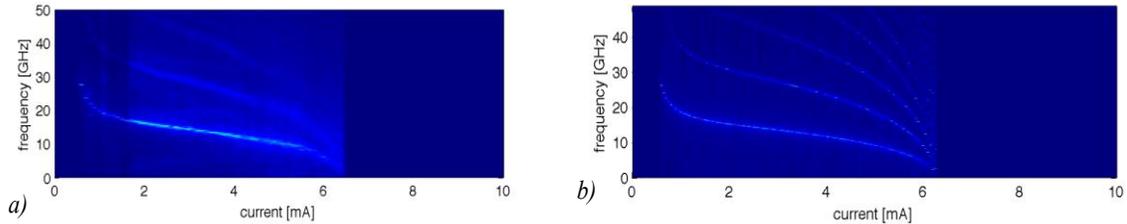

**Figure 14. Frequency-current diagram for a 30 nm diameter STO in CoNi film. Panel a) comes from a full micromagnetic simulation and b) from a macrospin model. Material parameters are identical in the simulations and similar to STOs studied in [19]. The threshold and cut-off currents and the oscillation frequencies for all harmonics are quite well approximated by the single-domain model.**

This means that typically for $d < 50$ nm diameter STOs, there is possibility to develop lumped, ODE-based models [27].

## B. Associative properties of electrically coupled STOs

As a case study, we investigate a model of electrically coupled STOs. The magnetization oscillations modulate the STO resistance and the oscillation signal can be picked up and superposed to the driving current of each individual STO. A circuit schematic is shown in Figure 15 - this is an implementation of the star-architecture interconnection. This circuit can be straightforwardly modeled in the single-domain approximation and details about the model are given in [19]. The coupling strength of the STOs is determined by the GMR ratio of the STOs (i.e. how strongly their resistance is changing upon oscillations) and by the transconductance of the active amplifier interconnecting them. Typical giant-magnetoresistance (GMR)-based STOs deliver a few-ten microvolts of output voltage ($V_i^{out}$), which is picked up, summed with signals from all other oscillators, and broadcasted to the input of each oscillator ($i^{control}$). The broadcasted current is typically in the order of 0.1 mA on top of few milliamperes STO driving current.

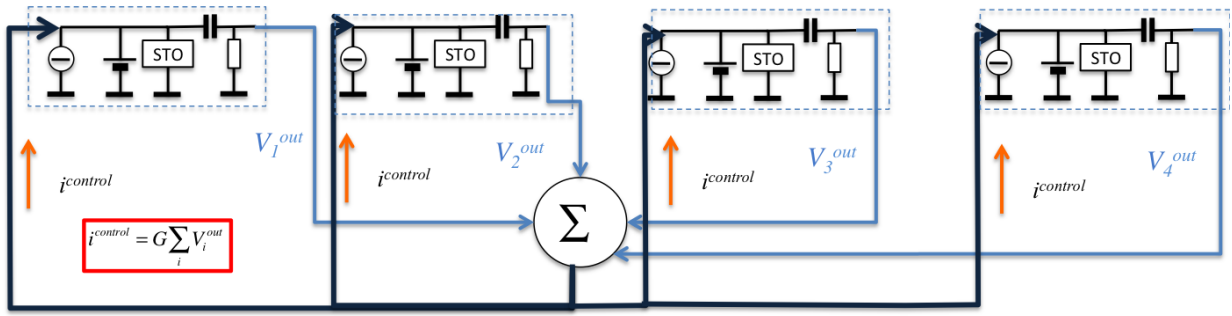

**Figure 15. A circuit schematics showing the interconnection of four STOs in the broadcast scheme (star architecture).**

We found that the phase of interconnected STOs is not robust and frequency shift keying (FSK) seems to be a more appropriate scheme for associative functions. Figure 16 shows the circuit dynamics for 64 STOs interconnected using the above scheme. We plot the instantaneous frequency (the inverse of the time elapsed between two zero crossings of the STO signal). The coupling is abruptly switched on at $t=5$ ns. In Figure 16a, the uncoupled oscillator frequencies are evenly spaced and lie too far apart to synchronize – pair-wise synchronization occurs between particular frequencies, but no dominant frequency component emerges. If some oscillator frequencies form a group of like frequencies (as shown in Figure 16b, then all these oscillators synchronize to a single dominant component. This frequency is not absolutely stable; it is slightly modulated by the group of unsynchronized oscillators. The strength and the stability of the dominant frequency component indicate the degree of matching.

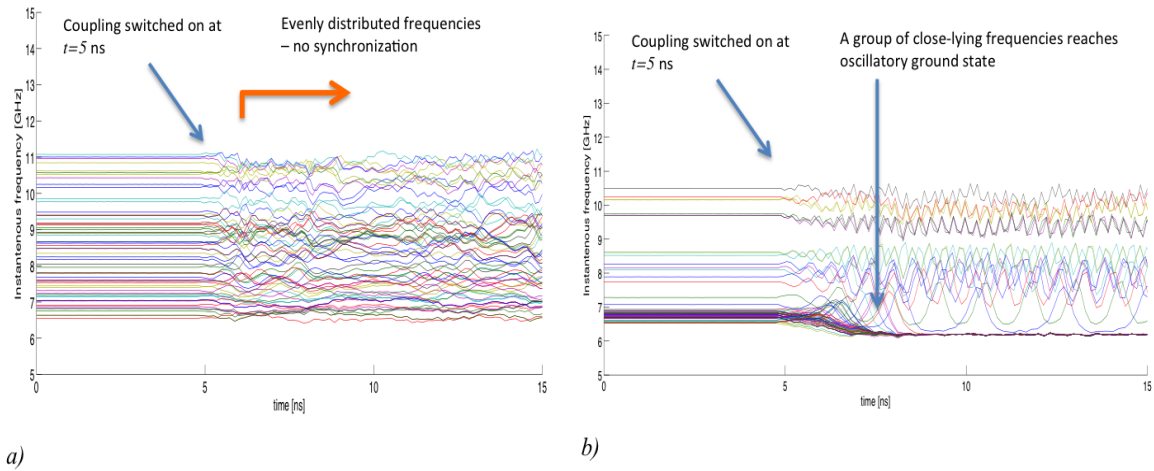

*a)* *b)*

**Figure 16. Oscillator frequencies are plotted as a function of time for 64 electrically interconnected STOs. In a) the oscillators frequencies are lying too far apart to synchronize, b) shows a group of STOs frequency-locking.**

## 6. Recognition accuracy and information gain

In this section, we examine the performance of several associative memory models. We also demonstrate that an associative memory that employs sparse distributed representations can perform pattern retrieval, and ultimately object recognition using real world visual application data.

John Hopfield postulated a theory of how the brain might work based on spin-glass dynamics which are well characterized mathematically [28]. He proposed a fully connected neural network initialized to some arbitrary state, and allowed to operate until it stabilizes. The energy of the network is defined:

$$E = -\frac{1}{2}\sum_{i,j} w(i,j)x(i)x(j) \qquad (18)$$

where $w(i,j)$ represents the connection strength between neuron $i$ and neuron $j$, and $x$ is the input pattern. Also, the weights are symmetrical with $w(i,j) = w(j,i)$. During operation, the network constantly tries to minimize total energy eventually terminating at a final point attractor. Attractors, therefore, sit in basins of low energy with the network state always trying to go "downhill". In general, attractor networks can be thought of as de-noising filters and can have multiple attractors, along with "repellers" and "saddles" which are meta-stable states.

When viewed from an information theoretic stance, attractor networks can be used to approximate Bayesian inference [29]. The idea is to shape the energy surface of a given network in such a way that it approximates the ideal energy surface that performs the required posterior calculation.

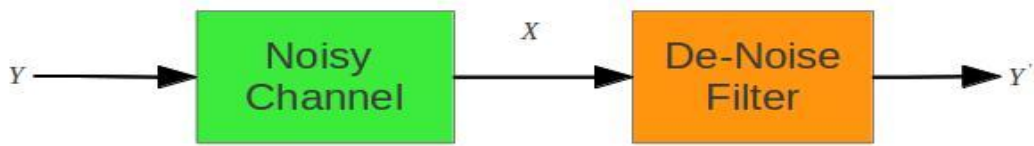

**Figure 17: A simple communication model.**

For example, consider the communication model illustrated in Figure 17. Here, the de-noising filter is modeled using an attractor network. A set of $L$ original messages, $Y = \{y_1, y_2, y_3 \ldots y_L\}$ are transmitted via a noisy communication channel producing a set of intermediate messages $X = \{x_1, x_2, \ldots, x_L\}$. These are passed through the denoising filter yielding a set of received messages $Y' = \{y'_1, y'_2, y'_3 \ldots y'_L\}$.

Now suppose we receive some new message $x$. The inference problem is then to find the index $l$ of the most likely original message within the original set $Y$,

$$l = \underset{Y}{argmax}\, P(Y|x) \tag{19}$$

where each posterior $P(y_i|x)$ is calculated using Bayes rule,

$$P(y_i|x) = \frac{P(x|y_i)P(y_i)}{\sum_j P(x|y_j)P(y_j)} \tag{20}$$

Then $y_l$ represents the most likely original message given the data we received, and our knowledge of the data statistics, channel errors and the messages being generated.

However, exact and even approximate Bayesian inference on an arbitrary network has been shown to be intractable [30]. In any case, approximate inference is still useful under the right circumstances, as we show in the coming example.

In practice, the Hopfield model has a number of issues that limit its usefulness. More practical models were developed by Gunther Palm [31] with David Willshaw [32] developing similar ideas. During training, the Palm/Willshaw model learns Hebbian-like correlations between input patterns and output patterns.

Retrieval is performed via best match association utilizing a voting scheme where the top $k$ winners are chosen, and all others are suppressed. This is sometimes called $k$ winner-take-all and is the most computationally expensive and powerful part of the model. It is widely known that in brain circuits, this function is performed via lateral inhibition within groups of neighboring neurons. In our implementation, we sort the voting results and use a threshold to calculate winners. However this method can produce ambiguous results when 2 or more votes are tied.

A related model is a type of sparse distributed memory developed by Furber [33]. This model uses two associative memories. The first, called an address decoder memory associates the input pattern to an intermediate pattern. The second called the data memory associates the intermediate pattern with the output pattern. With both the Palm and Furber models, best performance is achieved when the training patterns are sparse, meaning the number of 1s per $N$ bit pattern is on the order of $k = log_2(N)$. In addition, training patterns must be evenly distributed, meaning that they should have a minimum number of 1s in common.

In order to do useful computing, for example pattern recognition, the data constraints intrinsic to the Palm and Furber models require that real-world data first be mapped to a sparse distributed representation. Therefore, we developed a sparse coding step which employs dictionary pairs,

$$D = \{(u^{(i)}, v^{(i)}) \mid i = 1 \ldots L\} \tag{21}$$

where $u^{(i)}$ is a real-valued non-sparse (dense) centroid and $v^{(i)}$ is a sparse, binary, centroid. The real-valued dictionary is chosen from the input data using an unsupervised learning step (k-means). The sparse dictionary is chosen at random using sparse constraints.

To do sparse coding, the index $l$ of the closest dense centroid to the input pattern is used to index the sparse code book producing $v^{(l)}$. The sparse code $s^{(i)}$ is created by adding noise to the selected sparse centroid that is proportional to the Euclidian distance to the nearest real-valued centroid,

$$s^{(i)} = \Phi(v^{(l)}, \epsilon^{(i)}) \tag{22}$$

$$\epsilon^{(i)} = \alpha \lVert x^{(i)} - u^{(l)} \rVert \tag{23}$$

where $\Phi(v, \epsilon)$ moves $k\epsilon$ 1s in $v$ to other random locations within the vector and $\alpha$ is an adjustable parameter. This produces a sparse code which tends to preserve geometric information in the real-valued data. This is sometimes referred to similar input similar output (SISO), similarity producing coding (SPC), or similar input similar code (SISC) [34].

For the purposes of the examples shown in this text, we define Information Gain (IG) as the ability of an associative memory to recall memorized patterns. During the training process, training patterns are stored in the memory. Later, test patterns are used to recall the desired stored training patterns. Information gain is defined as,

$$IG = 1 - \frac{D_o}{D_i} \qquad (24)$$

where $D_i$ is the average Euclidean distance between input test and training patterns. Likewise, $D_o$ is the average distance between output test and training patterns.

Ideally the original pattern is recovered completely and $IG = 1$. Partial recovery then is some value greater than 0 but less than 1. If no signal is recovered $IG = 0$. Furthermore, if memory is severely dysfunctional, noise is added and $IG < 0$.

Figure 18 shows the information gain for the Hopfield, Palm and Furber models using random data. Note that in all three models, IG degrades as we increase the number of stored patterns. Likewise IG degrades as noise is added to the test data.

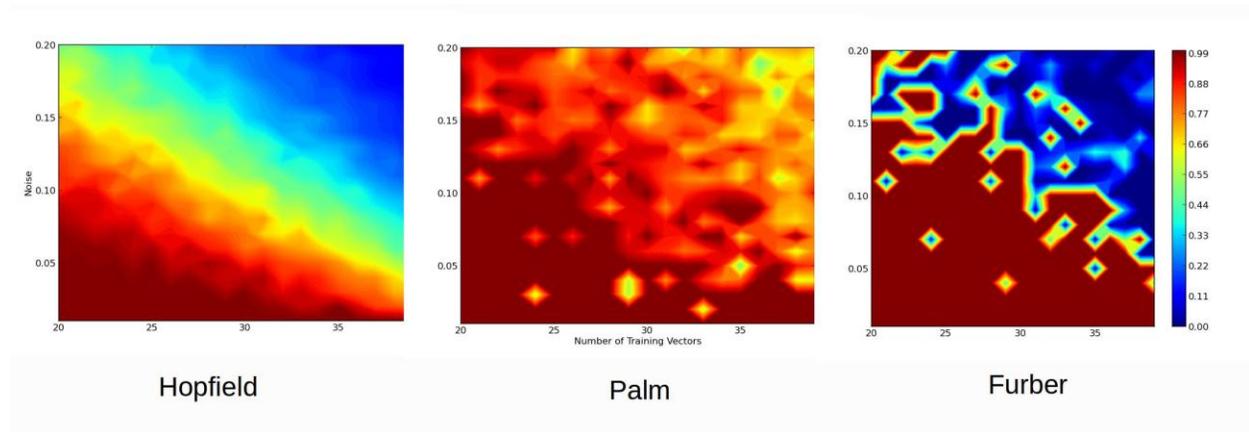

**Figure 18. Information gain plots for Hopfield, Palm and Furber networks.**

As a case study, we investigate using a Palm/Willshaw associative memory as part of a two-stage object recognition model. The first stage of the model processes raw pixel data producing HMAX descriptors [35]. The second stage of the model stores sparse distributed representations of the HMAX descriptors in an associative memory. Memorized patterns are later recalled during a visual object classification task using data collected from a mobile robot.

Our implementation of the HMAX algorithm contains four different hierarchical layers (S1: Gabor Filter Layer C1: Local invariance Layers S2: Intermediate Feature layers C2: Global invariance layers). Each layer processes information from the previous layer by performing a template matching or max pooling operation. In this way the model generates scale and position invariant image representations based on predefined features.

For our experiments, we created a visual application dataset using the Surveyor SRV-1 open source wireless mobile robot [36]. Image sets of four plastic cat-shaped figurines were taken at various distances and orientations with respect to the robot. The four classes of figurines are shown in Figure 19.

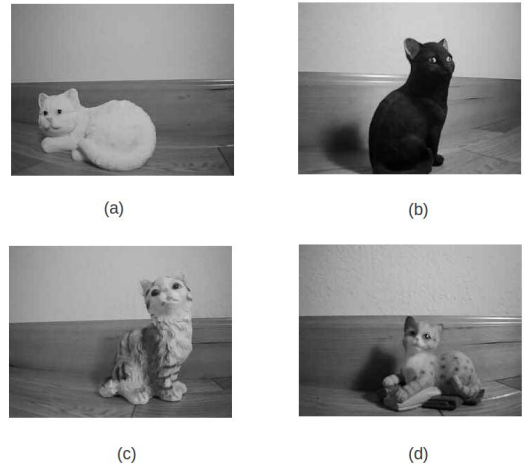

Figure 19. Four classes of figurines.

The robot was programmed to move forward a small amount toward a figurine, take a picture, and repeat, creating a set of 20 images. Using this method, we created six sets of images for each of the four classes of figurine. The six image sets were split into training and test data with 3 sets for training, and 3 for test.

We investigated object recognition results while varying both the input vector size, and adding increasing amounts of noise to the HMAX vector.

With notable exceptions [5] HMAX descriptors are real valued, high dimensionality vectors. Our implementation for instance produces a 8150 length output vector for each input image. We wanted to test the effect of compressing this large vector on the object recognition task. To do this, we used vector quantization.

To do vector quantization, the HMAX vector is first split into $p$ segments. Within each segment, and for each class, we create a codebook from the HMAX data with $q$ entries in each segment. During the quantization step, within each segment, we then calculate the average distance of the HMAX data for the segment to each entry in the codebook created for that segment. By varying the parameters for $p$ and $q$, we can create a compressed version of the original vector that has arbitrary length, while preserving as much information as possible from the original vector.

For each of the 20 different vector sizes, 20 different levels of noise were added to the training set by applying $\Phi(v, \epsilon)$ and varying $\epsilon$ to 20 different values. This created 400 unique sets of experimental parameters. For each set of parameters, accuracy, recall and precision were calculated over ten test rounds

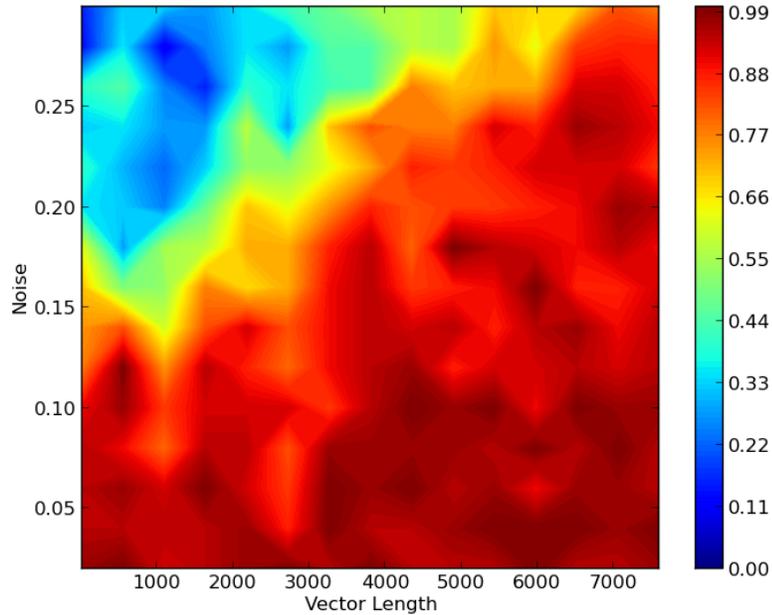

**Figure 20. HMAX+Palm object recognition accuracy.**

All the experiments used a 256 neuron Palm/Willshaw associative memory. We employed our SISO coding step, creating sparse descriptors which map the original real valued data into a sparse binary vector 256x1 with $k = log_2(256) = 8$ bits set. These descriptors were stored in a Palm/Willshaw associative memory model. During recall, HMAX data from the test images was used to recall the previously stored training descriptors. To do object recognition the hamming distance was computed between the associative memory output and the training set. The class with the smallest distance was counted as the winner. The accuracy results are shown in Figure 20. As expected, the model works better with the larger vector sizes because some information is lost when compressing the HMAX output vector to smaller sizes. Similarly, as noise is added to the HMAX descriptors, object recognition accuracy smoothly degrades.

## 7. Conclusions

In conclusion, we present a design of coupled oscillator associative memory array amenable to practical realization. We simulate its operation in both PSK and FSK schemes for realistic models of oscillators. The simulation shows successful recognition for the case of example patterns. We then expand our treatment to random patterns and patterns obtained from realistic images in order to estimate the accuracy of recognition.

## 8. Acknowledgements

The authors thank Steven Levitan, Tamas Roska, Matthew Puffal, and William Rippard for fruitful discussions. The authors are grateful for Intel Corporation support of this research.